\documentclass[aps,prd,onecolumn,groupedaddress,showpacs,nofootinbib,amssymb]{revtex4}
\usepackage{graphicx,bm}
\usepackage[english]{babel}
\usepackage{amsmath}
\usepackage{amssymb}
\usepackage{amsfonts}

\begin{document}

\def\cL{{\cal L}}
\def\be{\begin{equation}}
\def\ee{\end{equation}}
\def\bea{\begin{eqnarray}}
\def\eea{\end{eqnarray}}
\def\beq{\begin{eqnarray}}
\def\eeq{\end{eqnarray}}
\def\tr{{\rm tr}\, }
\def\nn{\nonumber \\}
\def\e{{\rm e}}


\title{Covariant power-counting renormalizable gravity: \\
Lorentz symmetry 
breaking and accelerating early-time FRW universe}

\author{Shin'ichi Nojiri$^{1,2}$
and Sergei D. Odintsov$^3$\footnote{Also at Tomsk State Pedagogical
University}}

\affiliation{ $^1$ Department of Physics, Nagoya University, Nagoya 464-8602,
Japan \\
$^2$ Kobayashi-Maskawa Institute for the Origin of Particles and
the Universe, Nagoya University, Nagoya 464-8602, Japan \\
$^3$Instituci\`{o} Catalana de Recerca i Estudis Avan\c{c}ats
(ICREA) and Institut de Ciencies de l'Espai (IEEC-CSIC), Campus
UAB, Facultat de Ciencies, Torre C5-Par-2a pl, E-08193 Bellaterra
(Barcelona), Spain}

\begin{abstract}

We continue the study of covariant power-counting renormalizable gravity
constrained by scalar Lagrange multiplier.
Lorentz symmetry breaking is investigated in such a theory in comparison 
with
the one in ghost condensation model. Covariant power-counting renormalizable
vector gravity which is invariant under $U(1)$ gauge symmetry is proposed.
Several forms of vector Lagrange multiplier in this theory are discussed.
It is shown that covariant scalar/vector gravity under consideration may
have power-law or de Sitter accelerating cosmological solution corresponding
to inflationary era.
Simplest black hole solution is obtained and dispersion relations
for graviton are presented.

\end{abstract}

\pacs{95.36.+x, 98.80.Cq}

\maketitle

\section{Introduction \label{I}}

The building of the consistent and satisfactory renormalizable theory of
quantum gravity is fundamental challenge for XXI century theoretical physics.
Despite the number of attempts (for the introduction, see book \cite{ilb})
this problem is still far from being solved.
The idea proposed in ref.\cite{Horava:2009uw} for renormalizable quantum
gravity
is to modify the ultraviolet
behavior of the graviton propagator in the Lorentz non-invariant way as
$1/\left|\bm{k}\right|^{2z}$, where $\bm{k}$ is the spatial momenta
and $z$ could be 2, 3 or larger integers.
They are defined by the scaling properties of space-time coordinates
$\left(\bm{x},t\right)$ as follows,
\be
\label{sym1}
\bm{x}\to b\bm{x}\, ,\quad t\to b^z t\, .
\ee
When $z=3$, the theory seems to be power-counting UV renormalizable.
In the construction of such theory, one introduces the
terms breaking the Lorentz invariance explicitly (or more precisely, breaking
full diffeomorphism invariance) by treating the temporal coordinate and the
spatial
coordinates in a different way.
Such model has the diffeomorphism invariance with respect only to the time
coordinate $t$ and spatial coordinates $\bm{x}$ transformations:
\be
\label{sym2}
\delta x^i=\zeta^i(t,\bm{x})\ ,\quad \delta t=f(t)\, .
\ee
Here $\zeta^i(t,\bm{x})$ and $f(t)$ are arbitrary functions.

In ref.\cite{Nojiri:2009th}, Ho\v{r}ava-like power-counting renormalizable
gravity with full
diffeomorphism invariance has been proposed.
When we consider the perturbations from the flat background, which has Lorentz
invariance, the Lorentz invariance of the propagator is dynamically broken by
the
non-standard coupling with a perfect fluid. The obtained propagator behaves as
$1/{\bm{k}}^{2z}$ with $z=2,3,\cdots$ in the ultraviolet region and the model
seems to
be perturbatively power-counting (super-)renormalizable if $z\geq 3$.
The price for such  renormalizability is the presence of mysterious fluid.
Recently \cite{Nojiri:2010tv}, this model has been rewritten in covariant form
in terms of the
scalar field constrained by the specific Lagrange multiplier. Hence, the model
of covariant
and power-counting renormalizable field theory of gravity
has been proposed. The effective fluid is induced
by imposing a constraint for a scalar field by using the Lagrange multiplier
field \cite{Lim:2010yk}.
Due to the constraint, the scalar field is not dynamical and even
in the high energy region, one can obtain a non-relativistic fluid.

In the present paper, we study the properties of covariant power-counting
renormalizable gravities and propose new versions of such theories.
The emergence of the early-time acceleration which may modify the inflationary
era is demonstrated. The paper is organized as follows.
In the next section we briefly review the covariant power-counting
renormalizable
gravity introduced in ref.\cite{Nojiri:2009th, Nojiri:2010tv}.
The Lagrange multiplier constraint and flat background expansion are discussed.
Section three is devoted to the detailed study of Lorentz symmetry breaking in
such a theory.
The comparison with Lorentz symmetry breaking in ghost condensation model is
done. In section four the covariant power-counting renormalizable vector
gravity is proposed. Its different versions are presented and symmetries as
well as different limits are investigated. Section five is devoted to the
construction of accelerating early-time FRW cosmology.
It is shown that time-dependent accelerating universe or de Sitter universe
occurs. The emergence of finite-time future singularity and its avoidance via
further modification of the theory under discussion is investigated.
Some outlook is given in the last section.
In the Appendix A the scalar-tensor presentation of covariant gravity is given
where higher derivative gravitational terms are absorbed by additional scalars.
Appendix B is devoted to the discussion of black hole solution and
dispersion relation for graviton.

\section{A model of covariant and power-counting renormalizable
field theory of gravity \label{II}}

In this section we review the construction of power-counting renormalizable,
covariant
gravity introduced in refs.\cite{Nojiri:2009th, Nojiri:2010tv}.
The action of the model is given by
\bea
\label{HrvHL12}
S_{2n+2} &=& \int d^4 x \sqrt{-g} \left\{ \frac{R}{2\kappa^2} - \alpha \left\{
\left(\partial^\mu \phi \partial^\nu \phi \nabla_\mu \nabla_\nu
+ 2 U_0 \nabla^\rho \nabla_\rho \right)^n
\left( \partial^\mu \phi \partial^\nu \phi R_{\mu\nu} + U_0 R \right)\right\}^2
\right. \nn
&& \left. - \lambda \left( \frac{1}{2} \partial_\mu \phi \partial^\mu \phi
+ U_0 \right) \right\}\, .
\eea
for $z=2n + 2$ model $\left( n=0,1,2,\cdots\right)$, and
\bea
\label{HrvHL13}
S_{2n+3} &=& \int d^4 x \sqrt{-g} \left\{ \frac{R}{2\kappa^2} - \alpha \left\{
\left(\partial^\mu \phi \partial^\nu \phi \nabla_\mu \nabla_\nu
+ 2 U_0 \nabla^\rho \nabla_\rho \right)^n
\left( \partial^\mu \phi \partial^\nu \phi R_{\mu\nu} + U_0 R \right)\right\}
\right. \nn
&& \left.
\times \left\{
\left(\partial^\mu \phi \partial^\nu \phi \nabla_\mu \nabla_\nu
+ 2 U_0 \nabla^\rho \nabla_\rho \right)^{n+1}
\left( \partial^\mu \phi \partial^\nu \phi R_{\mu\nu} + U_0 R \right)\right\}
   - \lambda \left( \frac{1}{2} \partial_\mu \phi \partial^\mu \phi
+ U_0 \right) \right\}\, .
\eea
for $z=2n + 3$ model $\left( n=0,1,2,\cdots\right)$ (compare with
\cite{carloni}).

As the generalization, the cosmological constant $\Lambda$ term may be added
to the action (\ref{HrvHL12}) and (\ref{HrvHL13}) as
\be
\label{sym43}
S = S_{2n+2} - \frac{\Lambda}{2\kappa^2}\int d^4 x \sqrt{-g} \, ,\quad
\mbox{or} \quad
S = S_{2n+3} - \frac{\Lambda}{2\kappa^2}\int d^4 x \sqrt{-g} \, .
\ee
The cosmological constant term does not change the ultraviolet structure of the
graviton propagator and
therefore the power-counting renormalizability.
Furthermore, in addition to the cosmological term, the kinetic scalar field
term may be added
\be
\label{sym44}
S = S_{2n+2} + \int d^4 x \sqrt{-g} \left( - \frac{\Lambda}{2\kappa^2}
   - \frac{1}{2} \partial_\mu \phi \partial^\mu \phi\right) \, ,\quad \mbox{or}
\quad
S = S_{2n+3} + \int d^4 x \sqrt{-g} \left( - \frac{\Lambda}{2\kappa^2}
   - \frac{1}{2} \partial_\mu \phi \partial^\mu \phi\right) \, .
\ee
Note, however, if we shift the Lagrange multiplier field as $\lambda\to \lambda
-1$,
the action (\ref{sym44}) is transformed as
\be
\label{sym50}
S = S_{2n+2} + \int d^4 x \sqrt{-g} \left( - \frac{\Lambda}{2\kappa^2} + U_0
\right)
\, ,\quad \mbox{or} \quad
S = S_{2n+3} + \int d^4 x \sqrt{-g} \left( - \frac{\Lambda}{2\kappa^2} + U_0
\right) \, .
\ee
Then as a result, the cosmological constant is shifted by
\be
\label{sym51}
\Lambda \to \Lambda - 2\kappa^2 U_0 \, .
\ee
Hence, the ultraviolet properties do not change
if the kinetic scalar field term is included.

Note that the above actions (\ref{HrvHL12}) and (\ref{HrvHL13}) (and
(\ref{sym43})
and (\ref{sym44})) are totally diffeomorphism invariant as they are given
in terms of the local fields.
In the actions (\ref{HrvHL12}) and (\ref{HrvHL13}) ( (\ref{sym43}) and
(\ref{sym44})),
$\lambda$ is the Lagrange multiplier field, which gives a constraint
\be
\label{LagHL2}
\frac{1}{2} \partial_\mu \phi \partial^\mu \phi
+ U_0 = 0\, ,
\ee
that is, the vector $(\partial_\mu \phi)$ is time-like.
At least locally, one can choose the direction of time to be parallel to
$(\partial_\mu \phi)$.
Then Eq. (\ref{LagHL2}) has the following form:
\be
\label{LagHL3}
\frac{1}{2} \left(\frac{d\phi}{dt}\right)^2 = U_0 \, .
\ee
Then the spacial region becomes a hypersurface where $\phi$ is a constant since
the hypersurface is orthogonal to the vector $(\partial_\mu \phi)$.

On the other hand, by the variation of $\phi$,
for example for $z=2$ $\left( n=0 \right)$ case in (\ref{HrvHL12}), we find
\be
\label{CRG_phi1}
0 = 4 \alpha \partial^\mu \left\{ \partial^\nu \phi R_{\mu\nu}
\left( \partial^\rho \phi \partial^\sigma \phi R_{\rho\sigma} + U_0 R \right)
\right\} + \partial^\mu \left( \lambda \partial_\mu \phi \right) \, .
\ee
For $z\geq 3$ ($n \geq 1$ case in (\ref{HrvHL12}) or $n\geq 0$ case in
(\ref{HrvHL13})),
one gets rather complicated equation.

Let us consider the perturbation from the flat background $g_{\mu\nu} =
\eta_{\mu\nu} + h_{\mu\nu}$.
Then the curvatures have the following form:
\be
\label{Hrv2}
R_{\mu\nu} = \frac{1}{2}\left[ \partial_\mu \partial^\rho h_{\nu\rho}
+ \partial_\nu \partial^\rho h_{\mu\rho} - \partial_\rho \partial^\rho
h_{\mu\nu}
   - \partial_\mu \partial_\nu \left( \eta^{\rho\sigma} h_{\rho\sigma}
\right)\right]\ ,\quad
R = \partial^\mu \partial^\nu h_{\mu\nu} - \partial_\rho \partial^\rho
\left( \eta^{\rho\sigma} h_{\rho\sigma} \right)\ .
\ee
The following gauge condition is chosen:
\be
\label{Hrv3}
h_{tt} = h_{ti} = h_{it} = 0\ .
\ee
Then the curvatures (\ref{Hrv3}) look as:
\bea
\label{Hrv4}
&& R_{tt} = - \frac{1}{2}\partial_t^2 \left(\delta^{ij} h_{ij} \right) \ ,\quad
R_{ij} = \frac{1}{2}\left\{ \partial_i \partial^k h_{jk} + \partial_j
\partial^k h_{ik}
+ \partial_t ^2 h_{ij} - \partial_k \partial^k h_{ij} \right\}\ ,\nn
&& R = \partial^i \partial^j h_{ij} + \partial_t^2 \left(\delta^{ij} h_{ij}
\right)
   - \partial_k \partial^k \left(\delta^{ij} h_{ij} \right) \, ,
\eea
and we find
\bea
\label{HrvHL14}
\partial^\mu \phi \partial^\nu \phi R_{\mu\nu} + U_0 R &=&
U_0 \left\{ \partial^i \partial^j h_{ij}
   - \partial_k \partial^k \left(\delta^{ij} h_{ij} \right) \right\}\, , \nn
\partial^\mu \phi \partial^\nu \phi \nabla_\mu \nabla_\nu
+ 2 U_0 \nabla^\rho \nabla_\rho &=&
2U_0 \partial_k \partial^k \, .
\eea
In the ultraviolet region, where $\bm{k}$ is large,
the propagator behaves as $1/\left| \bm{k} \right|^4$ for $z=2$ ($n=0$) case in
(\ref{HrvHL12}) and therefore the ultraviolet behavior is rendered.
For $z=3$ ($n=0$) case in (\ref{HrvHL13}),
the propagator behaves as $1/\left| \bm{k} \right|^6$ and therefore
the model becomes power-counting renormalizable.
For $z=2n +2$ ($n\geq 1$) case in (\ref{HrvHL12}) or $z=2n+3$ ($n\geq 1$) case
in (\ref{HrvHL13}),
the model becomes power-counting super-renormalizable.
The dispersion relation of the graviton is then given by
\be
\label{sym29}
\omega = \alpha c_0 k^z\, ,
\ee
in the high energy region.
Here $c_0$ is a constant, $\omega$ is the angular frequency corresponding to
the energy
and $k$ is the wave number corresponding to momentum.
If $\alpha<0$, the dispersion relation becomes inconsistent and therefore
$\alpha$ should be positive.


In this section, we only considered the perturbation from the flat background
which has the Lorentz symmetry.
In principle, any manifold is flat in a microscopic limit (which is a part of
the definition of ``manifold'') and the renormalizablity is the problem in
such a short distance region. In the microscopic limit, all the manifold can be
regarded
as flat. Thus, the renormalizability should be discussed only
in the flat background.

Note that any gravity theory does not respect global Lorentz symmetry
since the theory is formulated in curved space-time.
Only when we consider the perturbation from the flat background, which has a
global Lorentz symmetry, we can discuss about the Lorentz symmetry.
Then what we need is a model which admits a solution where the space-time
itself is
flat and therefore Lorentz invariant but a vector and/or tensor has a
non-trivial value, which breaks the Lorentz symmetry.
In this section, $(\partial_\mu \phi)$ is used as a vector.
More general case will be discussed below.

The above construction may be generalized for $D$ dimensions.
If $z=D-1$, the model becomes power-counting renormalizable and if $z>D-1$,
the model becomes super-renormalizable.
For example, for $D=11$, which may correspond to the M-theory, if $n=4$ in
(\ref{HrvHL13}) is chosen, the power-counting renormalizable theory emerges.


\section{Structure of the Lorentz symmetry breaking \label{III}}

Let us study Lorentz symmetry breaking for the actions
(\ref{HrvHL12}) and (\ref{HrvHL13}).
The actions are manifestly invariant under the full diffeomorphism invariance,
which is a local symmetry.
The action has a shift symmetry
\be
\label{sym3}
\phi \to \phi + \phi_0 \, .
\ee
Here $\phi_0$ is a constant.

One can confirm that the actions admit a flat vacuum solution. Indeed,
field equations are:
\be
\label{sym4}
0 = \frac{1}{2\kappa^2} \left( R_{\mu\nu} - \frac{1}{2} g_{\mu\nu} R \right)
+ G^\mathrm{higher}_{\mu\nu}
   - \frac{\lambda}{2} \partial_\mu \phi \partial_\nu \phi + \frac{1}{2}
g_{\mu\nu}
\left( \frac{1}{2} \partial_\rho \phi \partial^\rho \phi + U_0 \right)\, .
\ee
Here $G^\mathrm{higher}_{\mu\nu}$ comes from the higher derivative term (the
second term)
in the actions (\ref{HrvHL12}) and (\ref{HrvHL13}).
When we assume the flat vacuum solution, Eq.(\ref{LagHL2}) given by the
variation over $\lambda$
has a form (\ref{LagHL3}). Since for the flat vacuum solution all the
curvatures vanish,
Eq.(\ref{sym4}) reduces to
\be
\label{sym5}
0 = \lambda \partial_\mu \phi \partial_\nu \phi\, ,
\ee
whose solution is $\lambda=0$ since $\partial_\mu \phi\neq 0$ due to the
constraint
equation (\ref{LagHL2}) (if we choose the coordinate system properly, we have
$\partial_t \phi = \sqrt{2U_0}$ and $\partial_i \phi = 0$). Hence, the actions
(\ref{HrvHL12}), and (\ref{HrvHL13}) admit the flat vacuum solution
with $\lambda=0$.

We now consider the perturbation from the flat vacuum solution
$g_{\mu\nu} = \eta_{\mu\nu} + h_{\mu\nu}$ as in (\ref{Hrv2}).
Then the actions (\ref{HrvHL12}) and (\ref{HrvHL13})
obviously enjoy the Poincar\' e symmetry, that is, the actions are invariant
under the
Lorentz transformation and the translations. In the flat vacuum background,
a solution of (\ref{LagHL3}) is given by
\be
\label{sym6}
\phi = \sqrt{2U_0} t\, .
\ee
Since the solution depends on the time coordinate $t$, the solution
spontaneously
breaks the symmetry under the time translation:
\be
\label{sym7}
t \to t + t_0
\ee
Here $t_0$ is a constant. The solution (\ref{sym6}) also breaks
the shift symmetry in (\ref{sym3}).
We should note that the diagonal symmetry of the time
translation (\ref{sym7}) and the shift symmetry (\ref{sym3}) is not broken.
In fact, if we choose $\phi_0$ in (\ref{sym3}) as
\be
\label{sym8}
\phi_0 = - \sqrt{2U_0} t_0 \, ,
\ee
under the simultaneous transformation, the solution (\ref{sym6}) is invariant.
The diagonal symmetry effectively plays the role of the time translation and
the flat vacuum solution is effectively invariant under the time translation
(for rigorous discussion of Lorentz symmetry breaking in effective theories,
see \cite{ArmendarizPicon:2010mz}).
This structure is almost identical with that in the ghost condensation
model \cite{ArkaniHamed:2003uy},
which is a variation of the k-inflation \cite{ArmendarizPicon:1999rj} or
k-essence \cite{Chiba:1999ka}
models. The ghost sector of the ghost condensation model is given by
\be
\label{sym9}
S_\mathrm{ghost} = \int d^4 x \sqrt{-g } P \left(g^{\mu\nu}
\partial_\mu \phi \partial_\nu \phi\right)\, .
\ee
Here $P \left(g^{\mu\nu} \partial_\mu \phi \partial_\nu \phi\right)$ is
an appropriate function of
$g^{\mu\nu} \partial_\mu \phi \partial_\nu \phi$.
By the variation of $\phi$, one obtains
\be
\label{sym10}
0 = \partial_\mu \left( \sqrt{-g} g^{\mu\nu} \partial_\nu \phi
P' \left(g^{\mu\nu} \partial_\mu \phi \partial_\nu \phi\right) \right)\, .
\ee
In the flat background $g_{\mu\nu} = \eta_{\mu\nu}$, a solution of
(\ref{sym10}) is given by
\be
\label{sym11}
\phi = f_0 t\, .
\ee
Here $f_0$ is a constant. The solution (\ref{sym11}) is identical with that of
(\ref{sym6}) if
we identify $f_0 = \sqrt{2U_0}$.
Since the vector $\partial_\mu \phi$ has a non-vanishing value, the Lorentz
symmetry
is broken in both of the ghost condensation model and our models
(\ref{HrvHL12})
and (\ref{HrvHL13}).
There is, however, a big difference between the ghost condensation model and
our models.
When we consider the fluctuation from the solution (\ref{sym11}) in the ghost
condensation model $\phi = f_0 t + \delta\phi$, the fluctuation has a
propagating mode.
In our models, however, if we denote the fluctuation from the solution
(\ref{sym6}) as
$\phi = \sqrt{2U_0} t + \delta \phi$, Eq. (\ref{LagHL2}) shows that
\be
\label{sym12}
\frac{\partial \delta\phi}{\partial t} = 0\, ,
\ee
whose solution is given by
\be
\label{sym13}
\delta \phi = f\left(\bm{x} \right)\, .
\ee
Here $f\left(\bm{x} \right)$ is an arbitrary function of the spatial coordinate
$\bm{x}$.
Since $\delta\phi$ does not depend on the time coordinate, there is no
oscillating mode.
Due to the existence of the oscillating mode in the ghost condensation model,
the Lorentz
symmetry restores in the ultraviolet region and the breakdown occurs only in
the infrared
region. On the other hand, since there is no propagating mode of the scalar
field $\phi$
in our model, the Lorentz symmetry breaking occurs even in the ultraviolet
region.
Since we like to modify the propagator of the graviton in the ultraviolet
region, the symmetry
breaking should survive even in the ultraviolet region. This situation is very
different from
that in the usual spontaneous symmetry breaking.

In the ghost condensation model or our models, the Lorentz symmetry breaking is
spontaneous. The usual $U(1)$ Higgs model, whose potential is given by
\be
\label{sym14}
V_\mathrm{Higgs} = - \frac{m^2}{2} \phi^* \phi
+ \frac{\lambda^2}{4} \left(\phi^* \phi \right)^2\, ,
\ee
has a global $U(1)$ symmetry, which is the invariance under the transformation
\be
\label{sym15}
\phi \to \e^{i\theta_0} \phi\, ,
\ee
with a constant real parameter $\theta_0$.
In (\ref{sym14}), $\phi$ is a complex scalar field and $m$ and $\lambda$ are
positive parameters.
The minimum of the potential is given by
\be
\label{sym16}
\phi = \frac{\e^{i\varphi}m}{\lambda}\, .
\ee
Here $\varphi$ is a constant phase. The value of $\varphi$ can be arbitrary. If
one chooses specific
value of $\varphi$, the value of $\varphi$ is changed under the $U(1)$
transformation (\ref{sym15}) as
$\varphi \to \varphi + \theta$, and therefore the ground state is not invariant
under
the $U(1)$ transformation and the $U(1)$ symmetry breaks spontaneously.
One can always choose the real axis of the complex $\phi$-plane to be parallel
with the value of $\phi$
in the ground state so that $\varphi=0$.

In our model, the constraint equation (\ref{LagHL2}) shows that the value of
the vector $(\partial_\mu \phi)$
is located on the hyperboloid defined by
\be
\label{sym17}
   - x^\mu x_\mu \equiv t^2 - \bm{x}^2 = 2 U_0 \, .
\ee
The value of the vector $(\partial_\mu \phi)$ changes on the
hyperboloid under the Lorentz transformation. If we choose a value of
$(\partial_\mu \phi)$, the
Lorentz symmetry is broken spontaneously. After that we can always choose the
time axis
to be parallel to the vector $(\partial_\mu \phi)$.


Thus, there have been proposed several models, for example, ghost condensation
one,
where the Lorentz symmetry is broken spontaneously but
only in the infrared region. We need the Lorentz symmetry breaking in the
ultraviolet
region, where the renormalizability issue becomes a problem.
At least, the authors do not know such a model except the one in this paper,
where the Lorentz symmetry is spontaneously broken
even in the ultraviolet region.

Note that the Lorentz symmetry breaking is not directly
related with the renormalizablity. Only if the symmetry breaking improves the
ultraviolet
behavior of the propagator as in this paper, the renormalizability properties
are changed.
For vector theory with spontaneously broken Lorentz symmetry, as in the
next section, the explicit coupling improves the renormalization properties.

\section{Covariant vector gravity \label{IV}}

The spontaneous Lorentz symmetry breaking occurs when the quantity like vector
or tensor, which
is generally not invariant under the Lorentz transformation, has non-trivial
vacuum expectation value.
An exception occurs in case of the rank 2 symmetric tensor $B_{\mu\nu}$: if the
vacuum expectation
value of $B_{\mu\nu}$ is proportional to $\eta_{\mu\nu}$: $B_{\mu\nu} = c
\eta_{\mu\nu}$
with a constant $c$, the Lorentz symmetry does not break since $\eta_{\mu\nu}$
is invariant under
the Lorentz transformation.
Then instead of the actions (\ref{HrvHL12}) and (\ref{HrvHL13}), by using a
vector field
$A_\mu$, we may consider
\bea
\label{sym19}
S_{2n+2,A} &=& \int d^4 x \sqrt{-g} \left\{ \frac{R}{2\kappa^2} - \alpha
\left\{
\left( A^\mu A^\nu \nabla_\mu \nabla_\nu + 2 U_0 \nabla^\rho \nabla_\rho
\right)^n
\left( A^\mu A^\nu R_{\mu\nu} + U_0 R \right)\right\}^2 \right. \nn
&& \left. - \lambda \left( \frac{1}{2} A_\mu A^\mu + U_0 \right) \right\}\, ,
\eea
for $z = 2 n + 2$ $\left( n = 0,1,2,\cdots \right)$ and
\bea
\label{sym20}
S_{2n+3,A} &=& \int d^4 x \sqrt{-g} \left\{ \frac{R}{2\kappa^2} - \alpha
\left\{
\left( A^\mu A^\nu \nabla_\mu \nabla_\nu + 2 U_0 \nabla^\rho \nabla_\rho
\right)^n
\left( A^\mu A^\nu R_{\mu\nu} + U_0 R \right)\right\} \right. \nn
&& \left. \times \left\{
\left( A^\mu A^\nu \nabla_\mu \nabla_\nu + 2 U_0 \nabla^\rho \nabla_\rho
\right)^{n+1}
\left( A^\mu A^\nu R_{\mu\nu} + U_0 R \right)\right\}
   - \lambda \left( \frac{1}{2} A_\mu A^\mu + U_0 \right) \right\}\, ,
\eea
for $z = 2 n + 3$ $\left( n = 0,1,2,\cdots \right)$.
A possible problem is that $A_\mu$ can take a random value as long as it is
time-like
at different points of the space-time. This could be compared with the vector
$(\partial_\mu\phi)$,
which is almost parallel even at different points of the space-time.
At least if we start with the background where the value of the vector field is
parallel at different points in the space-time, which is also a solution, the
gravity seems to be power-counting renormalizable.
In fact, the constraint equation given by the variation of the Lagrange
multiplier field $\lambda$
is given by
\be
\label{sym21}
0 = \frac{1}{2} A_\mu A^\mu + U_0\, .
\ee
A solution of (\ref{sym21}) is given by
\be
\label{sym22}
A_0 = \sqrt{2U_0}\, , \ A_i = 0\, ,\ (i=1,2,3)\, ,
\ee
which breaks the Lorentz symmetry spontaneously. Then
\bea
\label{sym23}
A^\mu A^\nu R_{\mu\nu} + U_0 R &=&
U_0 \left\{ \partial^i \partial^j h_{ij}
   - \partial_k \partial^k \left(\delta^{ij} h_{ij} \right) \right\}\, , \nn
A^\mu A^\nu \nabla_\mu \nabla_\nu
+ 2 U_0 \nabla^\rho \nabla_\rho &=&
2U_0 \partial_k \partial^k \, ,
\eea
which gives expressions identical with (\ref{HrvHL14}) and therefore we obtain
(super-)renormalizable theories for $z=3$ and $z=2n+2$.

In general, one can add the kinetic term for the vector field and require the
$U(1)$
gauge symmetry. Then the following model may be proposed:
\bea
\label{sym19kin}
S &=& \int d^4 x \sqrt{-g} \left\{ \frac{R}{2\kappa^2} - \alpha \left\{
\left( \left( A^\mu - \partial^\mu \varphi \right) \left(A^\nu - \partial^\nu
\varphi \right)
\nabla_\mu \nabla_\nu + 2 U_0 \nabla^\rho \nabla_\rho \right)^n
\left( \left( A^\mu - \partial^\mu \varphi \right) \left(A^\nu - \partial^\nu
\varphi \right) R_{\mu\nu}
+ U_0 R \right)\right\}^2 \right. \nn
&& \left. - \frac{1}{4e^2} \left(\partial_\mu A_\nu - \partial_\nu A_\mu
\right)
\left(\partial^\mu A^\nu - \partial^\nu A^\mu \right)
   - \lambda \left( \frac{1}{2} \left( A_\mu - \partial_\mu \varphi \right)
\left(A^\mu - \partial^\mu \varphi \right)
+ U_0 \right) \right\}\, ,
\eea
for $z = 2 n + 2$ $\left( n = 0,1,2,\cdots \right)$ and
\bea
\label{sym20kin}
S &=& \int d^4 x \sqrt{-g} \left\{ \frac{R}{2\kappa^2} - \alpha \left\{
\left( \left( A^\mu - \partial^\mu \varphi \right) \left(A^\nu - \partial^\nu
\varphi \right) \nabla_\mu \nabla_\nu
+ 2 U_0 \nabla^\rho \nabla_\rho \right)^n
\left( \left( A^\mu - \partial^\mu \varphi \right) \left(A^\nu - \partial^\nu
\varphi \right) R_{\mu\nu}
+ U_0 R \right)\right\} \right. \nn
&& \times \left\{
\left( \left( A^\mu - \partial^\mu \varphi \right) \left(A^\nu - \partial^\nu
\varphi \right)
\nabla_\mu \nabla_\nu + 2 U_0 \nabla^\rho \nabla_\rho \right)^{n+1}
\left( \left( A^\mu - \partial^\mu \varphi \right) \left(A^\nu - \partial^\nu
\varphi \right) R_{\mu\nu}
+ U_0 R \right)\right\} \nn
&& \left. - \frac{1}{4e^2} \left(\partial_\mu A_\nu - \partial_\nu A_\mu
\right)
\left(\partial^\mu A^\nu - \partial^\nu A^\mu \right)
   - \lambda \left( \frac{1}{2} \left( A_\mu - \partial_\mu \varphi \right)
\left(A^\mu - \partial^\mu \varphi \right) + U_0 \right) \right\}\, ,
\eea
for $z = 2 n + 3$ $\left( n = 0,1,2,\cdots \right)$.
Here $e$ is the gauge coupling constant and
the scalar field $\varphi$ is the St\" uckelberg field
The actions (\ref{sym19kin}) and (\ref{sym20kin}) are invariant under the
$U(1)$ gauge transformation
\be
\label{U1a}
A_\mu \to A_\mu + \partial_\mu \epsilon\, ,\quad \varphi \to \varphi +
\epsilon\, .
\ee
Here $\epsilon$ is the (local) parameter of the the $U(1)$ gauge
transformation.
Especially if one chooses unitary gauge
\be
\label{U1b}
\varphi = 0\, .
\ee
the actions (\ref{sym19kin}) and (\ref{sym20kin}) reduce to
\bea
\label{sym19gf}
S &=& \int d^4 x \sqrt{-g} \left\{ \frac{R}{2\kappa^2} - \alpha \left\{
\left( A^\mu A^\nu \nabla_\mu \nabla_\nu + 2 U_0 \nabla^\rho \nabla_\rho
\right)^n
\left( A^\mu A^\nu R_{\mu\nu} + U_0 R \right)\right\}^2 \right. \nn
&& \left. - \frac{1}{4e^2} \left(\partial_\mu A_\nu - \partial_\nu A_\mu
\right)
\left(\partial^\mu A^\nu - \partial^\nu A^\mu \right)
   - \lambda \left( \frac{1}{2} A_\mu A^\mu + U_0 \right) \right\}\, ,
\eea
for $z = 2 n + 2$ $\left( n = 0,1,2,\cdots \right)$ and
\bea
\label{sym20gf}
S &=& \int d^4 x \sqrt{-g} \left\{ \frac{R}{2\kappa^2} - \alpha \left\{
\left( A^\mu A^\nu \nabla_\mu \nabla_\nu + 2 U_0 \nabla^\rho \nabla_\rho
\right)^n
\left( A^\mu A^\nu R_{\mu\nu} + U_0 R \right)\right\} \right. \nn
&& \times \left\{
\left( A^\mu A^\nu \nabla_\mu \nabla_\nu + 2 U_0 \nabla^\rho \nabla_\rho
\right)^{n+1}
\left( A^\mu A^\nu R_{\mu\nu} + U_0 R \right)\right\} \nn
&& \left. - \frac{1}{4e^2} \left(\partial_\mu A_\nu - \partial_\nu A_\mu
\right)
\left(\partial^\mu A^\nu - \partial^\nu A^\mu \right)
   - \lambda \left( \frac{1}{2} A_\mu A^\mu + U_0 \right) \right\}\, ,
\eea
for $z = 2 n + 3$ $\left( n = 0,1,2,\cdots \right)$.
For the solution (\ref{sym22}) of the constraint equation (\ref{sym21}),
the field strength $\partial_\mu A_\nu - \partial_\nu A_\mu$ vanishes and
therefore the corresponding energy-momentum tensor vanishes.
This shows that the flat space-time is a solution of the theory
(\ref{sym19kin}) and (\ref{sym20kin}).
We should note, however, there appears a constraint given by the variation of
$\varphi$, which
corresponds to the Gauss law constraint in QED. Due to the constraint, the
models (\ref{sym19kin})
and (\ref{sym20kin}) are different from the models (\ref{sym19}) and
(\ref{sym20}).

For the gauge fixed action (\ref{sym19gf}) and (\ref{sym20gf}) in the
ultraviolet region,
since we have the solution (\ref{sym22}),
the propagator behaves as $1/\left| \bm{k} \right|^4$ for $z=2$ ($n=0$) case in
(\ref{sym19gf}) and therefore the ultraviolet behavior is rendered.
For $z=3$ ($n=0$) case in (\ref{sym20gf}),
the propagator behaves as $1/\left| \bm{k} \right|^6$ and therefore
the model becomes power-counting renormalizable.
For $z=2n +2$ ($n\geq 1$) case in (\ref{sym19gf}) or $z=2n+3$ ($n\geq 1$) case
in (\ref{sym20gf}),
the model becomes power-counting super-renormalizable.

One may consider the ``weak'' coupling limit where $e\to 0$. Then in order for
the action
to be finite, the field strength should vanish: $\partial_\mu A_\nu -
\partial_\nu A_\mu$, which
indicates that the gauge field is the pure gauge. Therefore, the gauge field
can be rewritten as
$A_\mu = \partial_\mu \phi$. By substituting this expression into
(\ref{sym19gf}) and (\ref{sym20gf}),
the actions (\ref{HrvHL12}) and (\ref{HrvHL13}) are re-obtained.
On the other hand, one may consider the ``strong'' coupling limit where $e\to
\infty$. In the limit,
the kinetic term of the gauge field vanishes and the actions
in (\ref{sym19}) and (\ref{sym20}) are reprouced.
Note that the renormalizability does not depend on the magnitude of the
coupling.

Now, the role of constraint (\ref{sym21}) may be investigated. Instead of using
the Lagrange multiplier
field, the constraint can be realized by considering the following action:
\be
\label{Mass1}
S_{\lambda_0} = - \frac{\lambda_0}{2} \int d^4 x \sqrt{-g}
\left( \frac{1}{2} A_\mu A^\mu + U_0 \right)^2\, .
\ee
Here $\lambda_0$ is a constant. In the limit of $\lambda_0\to \infty$, the
constraint
(\ref{sym21}) follows. The mass $m_A$ of the vector field $A_\mu$ is given by
\be
\label{Mass2}
m_A^2 = \lambda_0 U_0\, .
\ee
Instead of the limit of $\lambda_0\to \infty$, it could be enough to choose
$\lambda_0$ so that
$m_A$ could be a cutoff scale for the renormalization.
In case of the string theory, the natural cutoff scale is the Planck scale. We
should also
note that there appears an infinite tower of the particle modes whose masses
are of the Planck scale order.
Then in the string theory, the cutoff scale mass appears always.
In the original proposal by Ho\v{r}ava \cite{Horava:2009uw}, the gravity model
was expected to be
an effective theory from the string theory.
Then the cutoff scale mass presence looks quite natural.

In principle, the constraint (\ref{sym21}) should be imposed only
in the high energy region for the graviton.
Then we may consider the following term:
\be
\label{Mass3}
S_{\lambda_G} = - \frac{\lambda_G}{2} \int d^4 x \sqrt{-g} R^n
\left( \frac{1}{2} A_\mu A^\mu + U_0 \right)^2\, .
\ee
Here $\lambda_G$ and $n$ are positive constant. When graviton has high energy,
$R$ becomes large,
and therefore the constraint (\ref{sym21}) appears in the high energy region.
It is interesting to remark that Lagrange multiplier constraint which 
breaks Lorentz invariance in the analogy with massive gravity (for recent 
review, see \cite{valery}) with Lorentz-symmetry breaking masses maybe 
also proposed.

One may also consider the following model:
\be
\label{sym24}
S = S_{2n+2} + \int d^4 x \sqrt{-g} F(G) \, ,
\ee
for $z=2n + 2$ $\left( n = 0,1,2,\cdots \right)$ model and
\be
\label{sym25}
S = S_{2n + 3} + \int d^4 x \sqrt{-g} F(G) \, ,
\ee
for $z=2n + 3$ $\left( n = 0,1,2,\cdots \right)$ model.
Here $G$ is the Gauss-Bonnet invariant defined by
\be
\label{sym27}
G = R^2 - 4 R_{\mu\nu} R^{\mu\nu} + R_{\mu\nu\rho\sigma} R^{\mu\nu\rho\sigma}
\, .
\ee
The $F(G)$-term in (\ref{sym24}) and (\ref{sym25}) does not change the
ultraviolet structure of the propagator of the graviton and therefore the
models
(\ref{sym24}) and (\ref{sym25}) remain to be power-counting renormalizable.
Thus, the construction of number of power-counting renormalizable vector
gravities is explicitly presented.

\section{Accelerating FRW cosmology \label{V}}

The gravitational terms different from general relativity in (\ref{HrvHL12})
and (\ref{HrvHL13}) are relevant
in the high energy region. Such terms might affect the inflationary era.
In this section, we briefly study FRW cosmology in the theory under discussion.
In order to obtain the FRW equations, the following
form of the metric is assumed:
\be
\label{bFRW}
ds^2 = - \e^{2b(t)}dt^2 + a(t)^2 \sum_{i=1,2,3} \left(dx^i\right)^2\, ,
\ee
and that the scalar field $\phi$ only depends on time. Then the
Eq.(\ref{LagHL2})
has the following form:
\be
\label{BBFRW}
\frac{1}{2} \left(\frac{d\phi}{dt}\right)^2 = \e^{2b(t)} U_0 \, .
\ee
Hence, one gets
\be
\label{HrvCos1}
\partial^\mu \phi \partial^\nu \phi R_{\mu\nu} + U_0 R = 6 U_0 \e^{-2b} H^2 \,
,\quad
\partial^\mu \phi \partial^\nu \phi \nabla_\mu \nabla_\nu
+ 2 U_0 \nabla^\rho \nabla_\rho = - 6 U_0 \e^{-2b} H \partial_t\, .
\ee
and the actions (\ref{HrvHL12}) and (\ref{HrvHL13}) have the following form:
\bea
\label{sym34}
S_{2n+2} &=& \int d^4 x a^3 \left[ \frac{\e^{-b}}{2\kappa^2} \left(6\dot H + 12
H^2 -
6\dot b H\right)
   - \left( 6 U_0 \right)^{2n+2} \e^b \left\{ \left( \e^{-2b} H \partial_t
\right)^n
\left( H^2 \e^{-2b} \right) \right\}^2 \right. \nn
&& \left. - \lambda \left( - \frac{\e^{-b}}{2}\left( \frac{d\phi}{dt} \right)^2
+ \e^b U_0 \right) \right]\, ,\\
\label{sym35}
S_{2n+3} &=& \int d^4 x a^3 \left[ \frac{\e^{-b}}{2\kappa^2} \left(6\dot H + 12
H^2 -
6\dot b H\right) \right. \nn
&& - 2^{2n+3}\cdot 3^2 \alpha\ U_0^{2n+2} \e^b \left\{ \left( \e^{-2b} H
\partial_t \right)^n
\left( H^2 \e^{-2b} \right) \right\}
\left\{ \left( \e^{-2b} H \partial_t \right)^{n+1} \left( H^2 \e^{-2b} \right)
\right\} \nn
&& \left. - \lambda \left( - \frac{\e^{-b}}{2}\left( \frac{d\phi}{dt} \right)^2
+ \e^b U_0 \right) \right]\, .
\eea
The first FRW equation looks as
\bea
\label{sym36}
0 &=& \frac{3}{\kappa^2}H^2 - \left( 6 U_0 \right)^{2n+2} \alpha a^{-3}
\left[ a^3 \left( D^n \left(H^2\right) \right)^2 - 4 (-1)^n H^2 {\bar D}^n
\left( a^3 D^n \left(H^2\right) \right) \right. \nn
&& \left. -4 \sum_{k=0}^{n-1} \left( -1 \right)^k \left( {\bar D}^{n-k}
\left(H^2\right) \right)
\left( {\bar D}^k \left( a^3 D^n \left(H^2\right) \right) \right) \right]
   - 2\lambda U_0 - \rho_\mathrm{matter}\, ,
\eea
for (\ref{sym34}) and as
\bea
\label{sym37}
0 &=& \frac{3}{\kappa^2}H^2 - \left( 6 U_0 \right)^{2n+3} \alpha a^{-3}
\left[ a^3 \left( D^n \left(H^2\right) \right) \left( D^{n+1} \left(H^2\right)
\right)
   - 2 (-1)^n H^2 {\bar D}^n \left( a^3 D^{n+1} \left(H^2\right) \right) \right.
\nn
&& - 2 (-1)^{n+1} H^2 {\bar D}^{n+1} \left( a^3 D^n \left(H^2\right) \right)
   -2 \sum_{k=0}^{n-1} \left( -1 \right)^k \left( {\bar D}^{n-k} \left(H^2\right)
\right)
\left( {\bar D}^k \left( a^3 D^{n+1} \left(H^2\right) \right) \right) \nn
&& \left. -2 \sum_{k=0}^{n} \left( -1 \right)^k \left( {\bar D}^{n+1-k}
\left(H^2\right) \right)
\left( {\bar D}^k \left( a^3 D^n \left(H^2\right) \right) \right) \right]
   - 2\lambda U_0 - \rho_\mathrm{matter}\, ,
\eea
for (\ref{sym35}).
Here $\rho_\mathrm{matter}$ is matter energy-density.
We also have put $b=0$ after the variation over $b$, where the metric
(\ref{bFRW}) reduces to the standard FRW metric and the operations of $D$ and
$\bar D$ for a scalar
$\varphi$ are defined by
\be
\label{sym38}
D\varphi \equiv H \frac{d\varphi}{dt}\, ,\quad
\bar D\varphi \equiv \frac{d\left( H\varphi\right)}{dt}\, .
\ee
On the other hand, by the variation over $a$, we get
\bea
\label{sym39}
0 &=& \frac{1}{\kappa^2} \left( 2\dot H + 3 H^2 \right)
+ 2^{2n+2} 3^{2n+1} \alpha U_0^{2n+2} \left\{ - 3 \left(D^n \left(H^2\right)
\right)^2 + 4 \left(-1\right)^n a^{-3}\frac{d}{dt}\left( H {\bar D}^n
\left( a^3 {\bar D}^n \left(H^2\right) \right) \right) \right. \nn
&& \left. + 2\sum_{k=1}^n a^{-3}\frac{d}{dt}\left(\left(\frac{d}{dt}
\left( D^{n-k} \left(H^2\right)\right)\right)
\left( {\bar D}^{k-1}\left( a^3 D^n\left(H^2\right)\right) \right)\right)
\right\}
+ p_\mathrm{matter}\, ,
\eea
for (\ref{sym34}) and
\bea
\label{sym40}
0 &=& \frac{1}{\kappa^2} \left( 2\dot H + 3 H^2 \right)
+ 2^{2n+3} 3^{2n+2} \alpha U_0^{2n+2} \left\{ - 3 \left(D^n \left(H^2\right)
\right)\left(D^{n+1} \left(H^2\right) \right) \right. \nn
&& + 2 \left(-1\right)^n a^{-3}\frac{d}{dt}\left( H {\bar D}^n
\left( a^3 {\bar D}^{n+1} \left(H^2\right) \right) \right)
+ 2 \left(-1\right)^{n+1} a^{-3}\frac{d}{dt}\left( H {\bar D}^{n+1}
\left( a^3 {\bar D}^n \left(H^2\right) \right) \right) \nn
&& + \sum_{k=1}^n a^{-3}\frac{d}{dt}\left(\left(\frac{d}{dt}
\left( D^{n-k} \left(H^2\right)\right)\right)
\left( {\bar D}^{k-1}\left( a^3 D^{n+1}\left(H^2\right)\right) \right)\right)
\nn
&& \left. + \sum_{k=1}^{n+1} a^{-3}\frac{d}{dt}\left(\left(\frac{d}{dt}
\left( D^{n-k+1} \left(H^2\right)\right)\right)
\left( {\bar D}^{k-1}\left( a^3 D^n\left(H^2\right)\right) \right)\right)
\right\} + p_\mathrm{matter}\, ,
\eea
for (\ref{sym35}).
The simplest case is $n=0$ in (\ref{sym34}) when FRW equations are
\bea
\label{HrvHL16}
\frac{3}{\kappa^2} H^2 &=& - 108 \alpha U_0^2 H^4 + 2 \lambda U_0
+ \rho_\mathrm{matter} \, , \\
\label{HrvHL17}
   - \frac{1}{\kappa^2} \left( 2\dot H + 3 H^2 \right)
&=& 36 \alpha U_0^2 \left( 3H^4 + 4H^2 \dot H \right)
+ p_\mathrm{matter} \, .
\eea
At the early universe where the curvature was large, the contribution from
the Einstein term, which corresponds to the right-hand side in (\ref{HrvHL16})
and (\ref{HrvHL17}), and the contributions from the
matter $\rho_\mathrm{matter}$ and $p_\mathrm{matter}$, could be neglected.
Then a solution of (\ref{HrvHL17}) is given by
\be
\label{sym41}
H=\frac{4}{3t}\, ,
\ee
which expresses the (power law) accelerated universe expansion corresponding
to the one with perfect fluid with $w=-1/2$.
Eq.(\ref{HrvHL16}) gives
\be
\label{sym42}
\lambda = \frac{32 \alpha U_0}{3t^4}\, .
\ee
One may consider the actions (\ref{HrvHL12}) and (\ref{HrvHL13}), which contain
a cosmological constant $\Lambda$. Then the equations (\ref{HrvHL16})
and (\ref{HrvHL17}) look as
\bea
\label{Cosmo1}
\frac{3}{\kappa^2} H^2 &=& - 108 \alpha U_0^2 H^4 + 2 \lambda U_0
+ \frac{\Lambda}{2\kappa^2} + \rho_\mathrm{matter} \, , \\
\label{Cosmo2}
   - \frac{1}{\kappa^2} \left( 2\dot H + 3 H^2 \right)
&=& 36 \alpha U_0^2 \left( 3H^4 + 4H^2 \dot H \right) -
\frac{\Lambda}{2\kappa^2}
+ p_\mathrm{matter} \, .
\eea
In case that the contribution from the matter is neglected, there emerges
the de Sitter solution with constant $H=H_0$, Here $H_0$ is given by
solving the following algebraic equation:
\be
\label{Cosmo3}
0 = 108 \alpha U_0^2 H_0^4 + \frac{3}{\kappa^2} H_0^2 -
\frac{\Lambda}{2\kappa^2} \, ,
\ee
whose solution is given by
\be
\label{Cosmo4}
H_0^2 = - \frac{72}{\alpha U_0^2 \kappa^2}
+ \sqrt{ \left(\frac{72}{\alpha U_0^2 \kappa^2}\right)^2
+ \frac{\Lambda}{216 \alpha U_0^2 \kappa^2}}\, .
\ee
Hence, de Sitter universe which may correspond to inflationary era occurs
as the explicit solution. The study of cosmological perturbations here 
maybe done in the same way as in Ho\v{r}ava-Lifshitz gravity (see, for 
instance, ref.\cite{misao}).

In addition to the cosmological term, one may add the scalar field kinetic term
as
in (\ref{sym44}). Then the equations (\ref{HrvHL16}) and (\ref{HrvHL17}) are
modified as
\bea
\label{sym45}
\frac{3}{\kappa^2} H^2 &=& - 108 \alpha U_0^2 H^4 + 2 \lambda U_0
+ \frac{1}{2}\left( \frac{d\phi}{dt} \right)^2 + \frac{\Lambda}{2\kappa^2}
+ \rho_\mathrm{matter} \, , \\
\label{sym46}
   - \frac{1}{\kappa^2} \left( 2\dot H + 3 H^2 \right)
&=& 36 \alpha U_0^2 \left( 3H^4 + 4H^2 \dot H \right)
+ \frac{1}{2}\left( \frac{d\phi}{dt} \right)^2
   - \frac{\Lambda}{2\kappa^2} + p_\mathrm{matter} \, .
\eea
If the contribution from the matter is neglected, there appears
again de Sitter universe solutions
\be
\label{sym47}
H^2 = H_0^2 = - \frac{72}{\alpha U_0^2 \kappa^2}
+ \sqrt{ \left(\frac{72}{\alpha U_0^2 \kappa^2}\right)^2
+ \frac{\Lambda- 2\kappa^2 U_2}{216 \alpha U_0^2 \kappa^2}}\, .
\ee
if
\be
\label{sym48}
\Lambda > 4\kappa^2 U_2 \, .
\ee
Note that $\lambda$ is given by
\be
\label{sym49}
\lambda = - \frac{\left( \frac{d\phi}{dt} \right)^2}{4U_0} = - \frac{1}{2}\, .
\ee
We should note that the shift of the cosmological constant (\ref{sym51}),
corresponding to the shift of the Lagrange multiplier field $\lambda\to \lambda
-1$,
is consistent if we compare (\ref{sym47}) with (\ref{Cosmo4}).

In the presence of matter fluid with the equation of state (EoS) parameter $w$,
matter
energy-density $\rho_\mathrm{matter}$ behaves as
\be
\label{p1}
\rho_\mathrm{matter} \sim a^{-3(1+w)}\, .
\ee
Then if $w<-1$, the matter energy-density $\rho_\mathrm{matter}$ increases with
the
expansion of the universe.
In the Einstein gravity, when $w<-1$, the Hubble rate behaves as
\be
\label{p2}
H_\mathrm{Einstein} \sim \frac{- \frac{2}{3(1+w)}}{t_s - t}\, ,
\ee
since $H^2 \propto \rho_\mathrm{matter}$.
Here $t_s$ is a constant and therefore there appears a singularity at $t=t_s$,
which
is called ``Big Rip'' singularity.
In case of (\ref{Cosmo1}), near the singularity, the Hubble rate $H$ becomes
large and
$H^4$ term dominates and therefore we have $H^4 \propto \rho_\mathrm{matter}$,
whose solution
is given by
\be
\label{p3}
H \sim \frac{- \frac{4}{3(1+w)}}{t_s - t}\, .
\ee
Then there still appears a singularity at $t=t_s$.
Therefore the higher derivative term does not prevent the singularity
unlike to the case of modified $F(R)$ gravity where $R^2$-term cures
the singularity as it was observed in refs.\cite{sing}.

In order to avoid the singularity, one may add the following term to the
actions
(\ref{HrvHL12}) and (\ref{HrvHL13}):
\be
\label{p4}
S_{2n+2} \to S_{2n+2} + S_A\, ,\quad
S_{2n+3} \to S_{2n+3} + S_A\, ,\quad S_A \equiv - A_0 \int d^4 x \sqrt{-g}
\left( \partial^\mu \phi \partial^\nu \phi R_{\mu\nu} + U_0 R \right)^m\, .
\ee
In high energy region, the mass dimension of $d^3 x dt$ becomes $- 3 - z$. Then the operators
with the mass dimension less than or equal to $3+z$ are power-counting renormalizable. From the actions
(\ref{HrvHL12}) and (\ref{HrvHL13}), it is natural to assume the mass dimension of the scalar
field is $-1$. Then the dimension of the operator
$\left( \partial^\mu \phi \partial^\nu \phi R_{\mu\nu} + U_0 R \right)^m$ is
$2m$. Then we assume $2m \leq 3 + z = 5 + 2n$ for the action (\ref{HrvHL12})
or $2m<6+2n$ for the action (\ref{HrvHL13}).
As clear from (\ref{HrvHL14}), when we expand the term in the powers of
$h_{ij}$,
the series start with the $m$-th power of $h_{ij}$.
Therefore if $m\geq 3$, the term does not change the propagator and therefore
it does not affect the UV structure.
In the background (\ref{bFRW}), using (\ref{HrvCos1}) one gets
\be
\label{p5}
S_A = - A_0 \int d^4 x \left(6 U_0\right)^m \e^{\left(-2m + 1 \right)b}
H^{2m}\, .
\ee
Then in $n=0$ case, Eq. (\ref{HrvHL16}) becomes
\be
\label{p6}
\frac{3}{\kappa^2} H^2 = - 108 \alpha U_0^2 H^4
+ \left(2m - 1 \right) A_0 \left(6 U_0\right)^m H^{2m}
+ 2 \lambda U_0 + \rho_\mathrm{matter} \, ,
\ee
If there occurs a Big Rip type singularity, the second term of the r.h.s. and
the matter density $\rho_\mathrm{matter}$ would dominate and we would obtain
\be
\label{p7}
0 \sim \left(2m - 1 \right) A_0 \left(6 U_0\right)^m H^{2m} +
\rho_\mathrm{matter} \, .
\ee
However, the quantities in the r.h.s. are always positive if $A_0$ is positive,
which leads
to the inconsistency.
Therefore, the Big Rip singularity does not occur.
When curvature is large, if the first and second terms in the r.h.s. in
(\ref{p6}) dominate,
de Sitter space universe occurs. In the de Sitter space, the Hubble rate is
given by
\be
\label{p8}
H = \left( \frac{108 \alpha U_0^2}{\left(2m - 1 \right)
A_0 \left(6 U_0\right)^m } \right)^{\frac{1}{2m-4}}\, .
\ee
The above FRW cosmology is also realized for the theory with
the actions (\ref{sym19}) and (\ref{sym20}) where the vector field is included,
The solution of the constraint equation (\ref{sym21}) is given by,
instead of (\ref{sym22}),
\be
\label{sym22b}
A_0 = \e^{b(t)}\sqrt{2U_0}\, , \ A_i = 0\, ,\ (i=1,2,3)\, .
\ee
By using the solution (\ref{sym22b}), we obtain
(\ref{sym36}), (\ref{sym37}), (\ref{sym39}), and (\ref{sym40}), again.
Therefore, FRW cosmological solutions do not change from the cases of covariant
gravity with scalars. One may also add the cosmological term as,
\be
\label{sym52}
S = S_{2n+2,A} - \frac{\Lambda}{2\kappa^2}\int d^4 x \sqrt{-g} \, ,\quad
\mbox{or} \quad
S = S_{2n+3,A} - \frac{\Lambda}{2\kappa^2}\int d^4 x \sqrt{-g} \, .
\ee
Such extra term does not change the ultraviolet structure of the graviton
propagator
and therefore, power-counting renormalizability. The cosmological solutions are
again
identical to the ones obtained in this section.

Thus, we demonstrated that covariant power-counting renormalizable gravity
naturally
leads to accelerating early-time expansion which may correspond to inflationary
era.
However, the accelerating cosmology is the same for scalar or vector covariant
gravity under discussion.

\section{Discussion}

In summary, we constructed and investigated the covariant power-counting renormalizable
scalar and/or vector gravity which is constrained by scalar/vector Lagrange multiplier term.
The ultraviolet behavior of the theory is improved because of the Lorentz symmetry breaking
which occurs due to non-trivial coupling with the effective (scalar/vector) fluid.
The comparison of Lorentz symmetry breaking in the theory under consideration with
the same in ghost condensation model is done. The $U(1)$ gauge symmetry structure of
the vector covariant gravity is studied as well as weak and strong electromagnetic
coupling constant limits.
As proposed theory pretends to improve the ultraviolet behavior of gravity
at high energies, the early-time FRW cosmology is discussed.
We demonstrate that early-time inflation may occur in the theory under consideration.
It is described by power-law accelerated Hubble rate or by de Sitter universe behavior.
It is interesting that power-law accelerating FRW  evolution may end up at finite-time future singularity.
Nevertheless, the additional modification of the covariant action by extra higher
derivative term which does not destroy the good ultraviolet behavior of the graviton
propagator may cure the future singularity.
Some remarks about dispersion relations  compared with general relativity ones are also made.

The theory under discussion may be considered as some step towards to the construction of 
renormalizable quantum gravity. Indeed, the ultraviolet behavior of such theory is improved 
in the same sense as the one in Ho\v{r}ava-Lifshitz gravity where Lorentz symmetry is broken 
from the very beginning. In the covariant theory under consideration, Lorentz symmetry 
is broken dynamically as the result of the non-trivial coupling with the effective fluid. 
In order to understand better its renormalization properties, the one-loop renormalization 
should be done. This is quite non-trivial task due to presence of Lagrange multiplier. 
It will be considered elsewhere.

\section*{Acknowledgments \label{VI}}

We are indebted to T. Kugo for the discussion about the quantization of the
system with
constraints and V. Rubakov for the interest to this work.
This research has been supported in part
by MEC (Spain) project FIS2006-02842 and AGAUR(Catalonia) 2009SGR-994 (SDO),
by Global COE Program of Nagoya University (G07)
provided by the Ministry of Education, Culture, Sports, Science \& Technology
and by the JSPS Grant-in-Aid for Scientific Research (S) \# 22224003 (SN).

\appendix

\section{Scalar-tensor representation of covariant action \label{A}}

We now consider the structure of the higher derivative terms in the actions
(\ref{HrvHL12}) and (\ref{HrvHL13}).
First we should note that
\bea
\label{HD1}
&& \partial^\mu \phi \partial^\nu \phi R_{\mu\nu} + U_0 R =
\left( \partial^\mu \phi \partial^\nu \phi
   - \frac{1}{2}g^{\mu\nu} \partial_\rho \phi \partial^\rho \phi \right)
R_{\mu\nu}
= \left(R_{\mu\nu} - \frac{1}{2} g_{\mu\nu} R \right)\partial^\mu \phi
\partial^\nu \phi \, ,\nn
&& \partial^\mu \phi \partial^\nu \phi \nabla_\mu \nabla_\nu
+ 2 U_0 \nabla^\rho \nabla_\rho
= \left( \partial^\mu \phi \partial^\nu \phi
   - g^{\mu\nu} \partial_\rho \phi \partial^\rho \phi \right) \nabla_\mu
\nabla_\nu
= \partial^\mu \phi \partial^\nu \phi
\left(\nabla_\mu \nabla_\nu - g_{\mu\nu} \nabla^\rho \nabla_\rho \right)\, .
\eea
Then in case of $n=0$ in (\ref{HrvHL12}), by introducing a scalar field $\eta$,
one can rewrite the action as
\be
\label{HD2}
S_2 = \int d^4 x \sqrt{-g} \left\{ \frac{R}{2\kappa^2} + \alpha \left\{ \eta^2
   - 2 \eta \left(R_{\mu\nu} - \frac{1}{2} g_{\mu\nu} R \right)\partial^\mu \phi
\partial^\nu \phi \right\}
   - \lambda \left( \frac{1}{2} \partial_\mu \phi \partial^\mu \phi
+ U_0 \right) \right\}\, .
\ee
Note that the term like $\left(R_{\mu\nu} - \frac{1}{2} g_{\mu\nu} R
\right)\partial^\mu \phi \partial^\nu \phi$
appears as $\mathcal{O}(\alpha)$ correction in superstring theory.
For general $n$ in (\ref{HrvHL12}), the action may be rewritten in
scalar-tensor form with
$2n+1$ scalar fields $\eta$, $\xi_i$, and $\zeta_i$
$\left(i=1,2,\cdots,n\right)$:
\bea
\label{HD3}
S_2 &=& \int d^4 x \sqrt{-g} \left\{ \frac{R}{2\kappa^2} + \alpha \left\{
\eta^2
   - 2 \eta \partial^\mu \phi \partial^\nu \phi
\left(\nabla_\mu \nabla_\nu - g_{\mu\nu} \nabla^\rho \nabla_\rho \right)
\zeta_1 \right. \right. \nn
&& \left. + \sum_{i=1}^{n-1} \xi_i \left( \zeta_i - \partial^\mu \phi
\partial^\nu \phi
\left(\nabla_\mu \nabla_\nu - g_{\mu\nu} \nabla^\rho \nabla_\rho \right)
\zeta_{i+1} \right)
+ \xi_n \left( \zeta_n - \left(R_{\mu\nu}
   - \frac{1}{2} g_{\mu\nu} R \right)\partial^\mu \phi \partial^\nu \phi\right)
\right\} \nn
&& \left. - \lambda \left( \frac{1}{2} \partial_\mu \phi \partial^\mu \phi
+ U_0 \right) \right\}\, .
\eea
Similarly for (\ref{HrvHL12}), scalar-tensor action with $2n+2$ scalar fields
$\xi_i$ and $\zeta_i$
$\left(i=1,2,\cdots,n+1\right)$ looks as
\bea
\label{HD4}
S_2 &=& \int d^4 x \sqrt{-g} \left\{ \frac{R}{2\kappa^2} + \alpha \left\{
\zeta_1 \zeta_2
+ \sum_{i=1}^n \xi_i \left( \zeta_i - \partial^\mu \phi \partial^\nu \phi
\left(\nabla_\mu \nabla_\nu - g_{\mu\nu} \nabla^\rho \nabla_\rho \right)
\zeta_{i+1} \right) \right. \right. \nn
&& \left. \left. + \xi_{n+1} \left( \zeta_{n+1} - \left(R_{\mu\nu} -
\frac{1}{2} g_{\mu\nu} R \right)
\partial^\mu \phi \partial^\nu \phi\right) \right\}
   - \lambda \left( \frac{1}{2} \partial_\mu \phi \partial^\mu \phi
+ U_0 \right) \right\}\, .
\eea

\section{Black hole solution \label{B}}

The arguments around Eqs. (\ref{sym4}) and (\ref{sym5}) demonstrate that the
Schwarzschild black hole
\be
\label{sym28}
ds^2 = - \left( 1 - \frac{M}{r} \right) dt^2 + \left( 1 - \frac{M}{r}
\right)^{-1} dr^2
+ r^2 d\Omega^2\, ,
\ee
and Kerr black hole
\bea
\label{Kerr}
&& ds_4^2 = \Delta \tilde dt^2 - \frac{\Sigma^2}{\Delta} dr^2
   - \Sigma^2 d\theta^2 - \frac{\sin^2\theta}{\tilde A}\left(
d\varphi - \Omega dt\right) \nn
&& \tilde \equiv \frac{\Sigma^2\left(\Delta - a^2 \sin^2\theta\right)}
{\Sigma^4\Delta - 4a^2 M^2 r^2 \sin^2 \theta} \, ,
\quad \Omega \equiv \frac{2a M r \tilde A}{\Sigma^2} \, ,\quad
\Delta \equiv r^2 - 2 M r + a^2\ ,
\quad \Sigma^2 \equiv r^2 + a^2 \cos^2\theta\, ,
\eea
are solutions since the curvatures vanish.
In (\ref{sym28}), $d\Omega^2$ is the metric of two dimensional sphere with unit
radius.

For the Ho\v{r}ava gravity and the theories under consideration, the dispersion
relation of the graviton is given by (\ref{sym29}) in the high energy region.
Then the phase speed $v_\mathrm{p}$ and the group speed $v_\mathrm{g}$ are
given by
\be
\label{sym30}
v_\mathrm{p} = \frac{\omega}{k} = \alpha c_0 k^{z-1}\, ,\quad
v_\mathrm{g} = \frac{d\omega}{dk} = \alpha c_0 z k^{z-1}\, ,
\ee
which become larger and larger when $k$ becomes larger and goes beyond the
light speed.
This tells that even in (\ref{sym28}), the high energy graviton can escape from
the horizon.
Note that the horizon is null surface and therefore in the usual Einstein
gravity,
particle cannot escape from the horizon since the speed of the particle is
always less than or equal
to the light speed. In our model, however, the speed of the graviton can exceed
the light speed and
escape from the horizon.

The dispersion relations in (\ref{sym31}) and $k$-dependent speed of graviton
could give some effects
at the early universe. At the early universe, the quantum fluctuations with the
wave length
$\lambda = 1/2\pi k$ become classical ones when the wavelength is larger than
the horizon
radius $r_\mathrm{H}$:
\be
\label{sym31}
\lambda = \frac{1}{2\pi k} > r_\mathrm{H} \equiv \frac{c}{H}\, .
\ee
Here $H$ is the Hubble rate. In case of (\ref{sym31}), since the speed depends
on $k$,
the horizon radius depends on $k$ as
\be
\label{sym32}
r_\mathrm{H}(k) = \frac{v_\mathrm{g}}{H} = \frac{c_0 z k^{z-1}}{H}\, .
\ee
Then Eq. (\ref{sym31}) shows
\be
\label{sym33}
2\pi c_0 z k^z < H\, .
\ee
which tells that the fluctuations in the high frequency modes could be changed
from the usual Einstein gravity $2\pi k < H$ (we choose the parameter ``light
speed''
to be unity $c=1$).
The fluctuations could be observed via CMB. Thus, in future the possible
difference
in the high frequency region might be observed what may provide the additional
test in favor/against of theory under discussion.

\end{document}